\documentclass[%
twocolumn,
 amsmath,amssymb,
 aps, physrev,
]{revtex4-2}

\usepackage{graphicx}
\usepackage{dcolumn}
\usepackage{bm}
\usepackage{xcolor}

\begin{document}


\title{\textbf{Finite-size scaling in quasi-3D stick percolation} 
}%

\author{Ryan K. Daniels}
\affiliation{
 Department of Computer Science and Technology, \\ University of Cambridge
}
\email[]{rkd43@cam.ac.uk}%


\begin{abstract}
This work extends the universal finite-size scaling framework for continuum percolation from two-dimensional (2D) to quasi-three-dimensional (Q3D) stick systems, in which sequentially deposited wires of finite diameter stack vertically on a flat substrate. Using Monte Carlo simulation, the percolation threshold is determined for isotropic Q3D stick systems as $N_c l^2 = 6.850923 \pm 0.000014$, approximately $21.5\%$ above the established 2D value of $5.6373$. The threshold is shown to be independent of the wire diameter-to-length ratio $d/l$, reflecting the scale invariance of the contact topology under sequential deposition. Simulation results indicate that, as for 2D networks, the spanning probability of Q3D stick percolation on square systems with free boundary conditions collapses onto a single universal scaling function once a nonuniversal metric factor is introduced. This collapse is consistent with the universal scaling function established for 2D continuum and lattice percolation.
\end{abstract}

\maketitle
\newpage

\section{Introduction}

Random networks of metallic nanowires are used in a growing range of electronic and optoelectronic applications, from transparent conducting films for flexible displays and photovoltaics~\cite{Ye2014,Sannicolo2016,Li2020} to neuromorphic computing platforms, where self-assembled nanowire networks exhibit brain-like properties including reconfigurable dynamics, short- and long-term memory, and emergent criticality~\cite{Avizienis2012,DiazAlvarez2019,Milano2020,Manning2018}. For neuromorphic devices, operating near the percolation transition promotes the edge-of-criticality dynamics thought to underpin computational functionality~\cite{Stieg2012,Mallinson2019}. For transparent conductors, minimising density while maintaining conductivity is essential for optical transparency. In many of these applications, the network operates close to the percolation threshold, where the wire density is just sufficient to establish a connected path across the device. The percolation threshold and the finite-size scaling behaviour in its vicinity are therefore fundamental to device design.

For idealised two-dimensional stick systems, the percolation threshold is well established. Li and Zhang~\cite{LiZhang2009} determined the critical density for isotropic widthless sticks on square systems with free boundary conditions to high precision, $N_c l^2 = 5.63726 \pm 0.00002$, and showed that the spanning probability obeys the same universal finite-size scaling function as lattice percolation. However, real nanowire networks, especially those used for neuromorphic applications, are not two-dimensional. Wires deposited onto a substrate stack vertically: each new wire rests on those already present, and pairs that would cross in the plane may be separated in the vertical direction and fail to make electrical contact. Daniels and Brown~\cite{DanielsBrown2021} showed that this quasi-three-dimensional (Q3D) stacking produces networks with fundamentally different topologies --- lower mean degree, reduced clustering, diminished small-world character, and higher modularity. Crucially, the mean number of contacts per wire, which grows linearly with density in the 2D model, saturates to a constant in the Q3D case ~\cite{DanielsBrown2021, TarasevichEserkepov2026}. These topological differences are expected to affect emergent behaviour in nanowire networks~\cite{Daniels2022}, since properties such as reservoir computing performance and synchronisability in oscillator networks are sensitive to small-world connectivity~\cite{Loeffler2020,Deng2007}.

The contact saturation also has direct consequences for electrical transport. Tarasevich and Eserkepov~\cite{TarasevichEserkepov2026} recently showed, within a mean-field approximation, that the different contact scaling leads to qualitatively different conductivity behaviour: the electrical conductivity of a 2D network grows quadratically with wire density, whereas in the Q3D model the dependence is linear. The discrepancy is most severe when contact resistance dominates over wire resistance, where the 2D model can overestimate conductivity by up to two orders of magnitude.

Despite these advances, the most basic quantity characterising the Q3D system, its percolation threshold, has not been determined. An earlier estimate based on small system sizes without finite-size scaling extrapolation was reported in Ref.~\cite{DanielsMSc}, but the limited range of system sizes precluded a reliable determination of the percolation threshold. The topological studies of Daniels and Brown operated well above percolation and did not attempt a finite-size scaling analysis, while the conductivity theory of Tarasevich and Eserkepov assumes the system is already conducting. If the Q3D threshold differs from the 2D value, then device models based on 2D simulations predict both the wrong onset density for conduction and the wrong scaling of conductivity above that onset.

In this work, we address this gap by performing Monte Carlo simulations of stick percolation in a Q3D deposition model and extract the critical density using the finite-size scaling framework established by Li and Zhang. We first validate our methodology against the known 2D result, then extend it to the Q3D case and investigate the dependence of $N_c$ on the wire diameter-to-length ratio $d/l$.

\section{Finite-size scaling framework}
\label{sec:percolation}

Our goal is to determine the critical stick density $N_c$, the percolation threshold for both the 2D and Q3D models. To do so, we exploit the well-established finite-size scaling theory for percolation transitions~\cite{StaufferAharony2018, Privman1984}, as applied to continuum stick systems by Li and Zhang~\cite{LiZhang2009}.

\begin{table}
\caption{\label{tab:notation}Summary of the notation used throughout. The
fitted quantities ($N_c$, $a_1$, $a_3$, $a_5$, $b_0$, $K_3$, $K_5$) are
reported in Table~\ref{tab:results}.}
\begin{ruledtabular}
\begin{tabular}{ll}
Symbol & Description \\
\hline
$L$              & Linear system size                                   \\
$l$              & Stick (wire) length; set to unity                   \\
$d$              & Wire diameter (Q3D model)                            \\
$d/l$            & Diameter-to-length ratio                            \\
$N$              & Stick number density (per unit area)               \\
$N_c$            & Critical density (percolation threshold)          \\
$N_{0.5}(L)$     & Density at which $R=1/2$ for size $L$           \\
$R(N,L)$         & Spanning (crossing) probability                     \\
$R_c$            & Critical spanning probability, $R_c=1/2$            \\
$n_f$            & Stick count at first spanning                      \\
$\lambda$        & Poisson mean, $\lambda=NL^2$                      \\
$\nu$            & Correlation-length exponent, $\nu=4/3$            \\
$x$              & Scaling variable, $x=(N-N_c)L^{1/\nu}$              \\
$F(x)$           & Finite-size scaling function                       \\
$a_0,a_1,a_3,a_5$& Polynomial coefficients of $F(x)$; $a_0=1/2$        \\
$b_0$            & Leading finite-size correction amplitude          \\
$A$              & Nonuniversal metric factor, $A=a_1$                \\
$\hat{x}$        & Rescaled scaling variable, $\hat{x}=Ax$              \\
$\hat{F}(\hat{x})$& Universal finite-size scaling function (UFSSF)     \\
$K_3,K_5$        & Universal coefficients, $K_i=a_i/a_1^i$           \\
\end{tabular}
\end{ruledtabular}
\end{table}

Consider a square system of linear dimension $L$ containing sticks of unit length at number density $N$ sticks per unit area. The spanning probability $R(N, L)$ is the probability that a connected cluster of sticks bridges two opposite boundaries. In the vicinity of $N_c$, finite-size scaling theory~\cite{HoviAharony1996} gives
\begin{equation}
    R(N, L) \approx F(x) + \frac{b_0}{L},
    \label{eq:scaling}
\end{equation}
where $x = (N - N_c) L^{1/\nu}$ is the scaling variable, $\nu = 4/3$ is the two-dimensional correlation-length exponent, and $F(x)$ is a scaling function. The correction term $b_0/L$ accounts for the leading finite-size effect in systems with free (open) boundary conditions; it dominates over the subleading nonanalytic correction of order $L^{-\omega}$ (with $\omega \approx 0.9$) that arises from irrelevant scaling variables~\cite{HoviAharony1996, ZiffNewman2002}.

For small $x$, the scaling function admits a polynomial expansion
\begin{equation}
    F(x) = a_0 + a_1 x + a_3 x^3 + a_5 x^5 + \cdots,
    \label{eq:polynomial}
\end{equation}
in which only odd powers of $x$ appear beyond the constant term (i.e. $a_i = 0$ for even $i > 0$). This restriction follows from the self-duality of the square geometry under exchange of the spanning and non-spanning directions: the complementary probability $1 - R$ for the transverse direction must satisfy $1 - R(N_c, L) = R(N_c, L)$ in the $L \to \infty$ limit, fixing $a_0 = 1/2$ and requiring all even-order coefficients to vanish~\cite{Ziff1992, HoviAharony1996}. The critical spanning probability $R_c \equiv R(N_c, \infty) = 1/2$ is therefore universal for square systems with free boundaries.

In a Monte Carlo procedure (described in Section~\ref{sec:simulation}), each realisation deposits sticks one at a time until a spanning cluster forms, recording the total stick count $n_f$ at first spanning. The raw spanning probability $R_{n,L}$ is thus defined at integer stick counts $n$. To obtain $R(N, L)$ as a smooth function of the continuous density $N = n/L^2$, we convolve with the Poisson distribution~\cite{LiZhang2009}:
\begin{equation}
    R(N, L) = \sum_{n=0}^{\infty} \frac{\lambda^n e^{-\lambda}}{n!}\, R_{n,L},
    \label{eq:poisson}
\end{equation}
where $\lambda = N L^2$. This transforms the discrete, microcanonical spanning data into a grand-canonical ensemble at any desired density $N$ with arbitrary precision. The convolution is the continuum-percolation analogue of the binomial transform used in lattice percolation~\cite{NewmanZiff2000, ZiffNewman2002}; in that setting, the number of occupied bonds at a given occupancy $p$ is binomially distributed, whereas in a continuum system with a fixed expected density, the stick count follows a Poisson distribution.

With $R(N, L)$ for a range of system sizes $L$, the critical density is determined as follows. Define $N_{0.5}(L)$ as the density at which $R(N_{0.5}, L) = 1/2$. From Eq.~\eqref{eq:scaling}, this quantity converges to $N_c$ as
\begin{equation}
    N_{0.5}(L) - N_c = -\frac{b_0}{a_1}\, L^{-1-1/\nu} + \cdots,
    \label{eq:Nc_convergence}
\end{equation}
so that a plot of $N_{0.5}(L)$ against $L^{-1-1/\nu}$ should be linear for sufficiently large $L$, with the intercept giving $N_c$.

Throughout this work, we assume the two-dimensional correlation-length exponent $\nu = 4/3$. For the 2D stick system this is exact and well established. For the Q3D model, the assumption rests on the quasi-two-dimensional geometry of the system: although vertical stacking substantially alters the network topology relative to the freely-interpenetrating 2D model \cite{DanielsBrown2021}, the vertical extent of the network remains finite and does not scale with system size. A finite-thickness system is expected to remain in the 2D universality class, with the topological differences affecting nonuniversal quantities (the threshold and the metric factor) rather than the critical exponents.

Consistency of the extracted universal coefficients with the known 2D values, together with a successful collapse of the spanning-probability data, provides an a posteriori check on this assumption: a substantially incorrect $\nu$ should degrade the collapse and yield inconsistent coefficients. We emphasise, however, that this is a consistency check rather than an independent determination of $\nu$: the scaling variable is constructed using the assumed exponent, and the collapse quality is only weakly sensitive to moderate changes in $\nu$. This assume-and-verify approach follows established practice.

We extract the critical parameters using two complementary approaches. In the first, following Li and Zhang~\cite{LiZhang2009}, we proceed in two stages. The critical density is obtained by computing $N_{0.5}(L)$ for each system size and fitting the linear relation in Eq.~\eqref{eq:Nc_convergence} to extrapolate $N_c$ as the $L \to \infty$ intercept. With $N_c$ fixed to this value, the scaling function coefficients $\{a_i\}$ and the finite-size correction $b_0$ are then determined by fitting Eqs.~\eqref{eq:scaling}--\eqref{eq:polynomial} to the $R(N, L)$ data. The universal coefficients $K_i$ follow immediately from $K_i = a_i / a_1^i$. This two-step procedure has the advantage of transparency: the extrapolation plot provides a direct visual confirmation of the expected finite-size scaling behaviour, and any departures from linearity signal either corrections to scaling or insufficient system sizes.

In the second approach, we fit the polynomial model of Eqs.~\eqref{eq:scaling}--\eqref{eq:polynomial} directly to the $R(N, L)$ data across all system sizes simultaneously, treating $N_c$, $b_0$, and the coefficients $\{a_i\}$ as free parameters. This method is more statistically efficient, since it uses the full shape of $R(N, L)$ at each $L$ rather than compressing each curve to a single crossing point, and it avoids propagating the uncertainty in $N_c$ from the first stage into subsequent fits. Agreement between the two methods provides an internal consistency check on the extracted parameters.

A stronger test of the scaling framework is provided by the \emph{universal finite-size scaling function} (UFSSF). The scaling function $F(x)$ in Eq.~\eqref{eq:polynomial} is system-dependent through the coefficients $\{a_i\}$. However, upon introducing the nonuniversal metric factor $A$ and rescaling $\hat{x} = A\, x$, the resulting function
\begin{equation}
    \hat{F}(\hat{x}) = \tfrac{1}{2} + \hat{x} + K_3 \hat{x}^3 + K_5 \hat{x}^5 + \cdots
    \label{eq:UFSSF}
\end{equation}
is expected to be universal across all systems sharing the same dimensionality, boundary conditions, and aspect ratio~\cite{HoviAharony1996, HuLinChen1995}. This universality has been confirmed for a wide range of lattice models and was extended to continuum stick percolation by Li and Zhang, who found $K_3 = -1.055 \pm 0.002$ and $K_5 = 0.783 \pm 0.004$, in excellent agreement with lattice values~\cite{LiZhang2009, HoviAharony1996, HuLinChen1995}. Verifying that our simulation data collapse onto this same UFSSF serves as an independent check on both the extracted $N_c$ and the overall consistency of the scaling analysis for the 2D system. Furthermore, if the Q3D data collapse onto the same UFSSF with consistent $K_3$ and $K_5$ values, this provides evidence that the Q3D system is consistent with the same universality class as the 2D system.

\section{\label{sec:simulation}Simulation details}

The general simulation procedure in 2D is first described, and then the additional considerations for the Q3D case are addressed. We follow the approach of Li and Zhang~\cite{LiZhang2009} and use the same notation throughout.

\subsection{2D simulation}

We consider widthless sticks with uniformly distributed orientations, each of unit length $l = 1$, deposited on a square domain of side length $L$ with free (open) boundary conditions. For each stick, the midpoint $(c_x, c_y)$ is drawn uniformly from $[0, L]^2$ and the orientation angle $\theta$ is drawn uniformly from $[-\pi/2, \pi/2)$. The endpoints of the stick are then $(c_x \pm \tfrac{1}{2}\cos\theta,\; c_y \pm \tfrac{1}{2}\sin\theta)$.

The left and right boundaries of the domain serve as electrodes. Following Li and Zhang, each boundary is subdivided into $L$ unit-length segments that are pre-connected into a single cluster. This ensures that boundary segments are geometrically identical to deposited sticks for the purposes of intersection testing. The percolation condition is met when the left and right boundary clusters belong to the same connected component. Each realisation records $n_f$, the total number of deposited sticks at first spanning. The critical density $N_c$ is then extracted from an ensemble of such realisations across multiple system sizes $L$, using the fitting procedure described in Section~\ref{sec:percolation}. Note that the boundary sticks themselves are not included in the value of $n_f$. Two sticks belong to the same cluster if they intersect. The cluster connectivity structure is maintained using the weighted union-find algorithm with path halving~\cite{Tarjan1972, NewmanZiff2000, NewmanZiff2001}, adapted for continuum percolation by Li and Zhang~\cite{LiZhang2009}.




Early Monte Carlo studies of stick percolation~\cite{PikeSeager1974, BalbergBinenbaumAnderson1983, NedaFlorianBrechet1999} tested all pairs of sticks for intersection, resulting in $\mathcal{O}(n^2)$ scaling that limited simulations to small system sizes or few realisations. To avoid the $\mathcal{O}(n^2)$ cost of testing every pair of sticks for intersection, we employ the subcell decomposition~\cite{VicsekKertesz1981, WagnerBalbergKlein2006}, as applied to stick systems by Li and Zhang~\cite{LiZhang2009}.



\subsection{Extension to quasi-3D}

In the 2D model, sticks are treated as one-dimensional lines that interpenetrate freely: every crossing in the $xy$-plane registers as a contact. In a physical nanowire network, however, each wire has a finite diameter $d$ and occupies a volume that excludes other wires. Wires deposited sequentially onto a substrate therefore stack vertically, and a pair of wires that would cross in the 2D projection may be separated in the $z$-direction and fail to make contact. This quasi-three-dimensional (Q3D) stacking has been shown to produce networks with significantly different topological properties to the purely 2D case, including lower mean degree, reduced clustering, and diminished small-world character~\cite{DanielsBrown2021, Daniels2022}.

To investigate how stacking affects the percolation threshold, we extend the 2D simulation to a Q3D deposition model, following the framework of Daniels and Brown~\cite{DanielsBrown2021}. The $xy$-placement of each wire (midpoint, orientation) is identical to the 2D case.

Wires are deposited one at a time. For each new wire, the algorithm proceeds as follows. First, all crossings with previously deposited wires are identified in the $xy$-projection, using the same subcell grid as in the 2D case. At each projected crossing, the potential support height is computed: if the existing wire has centre-height $z_\mathrm{exist}$ at the crossing point, then the new wire's centre would sit at $z_\mathrm{exist} + d$ if resting on that support. A wire with no crossings lands flat on the substrate at height $z = d/2$.

\begin{figure*}
    \centering
    \includegraphics[width=\textwidth]{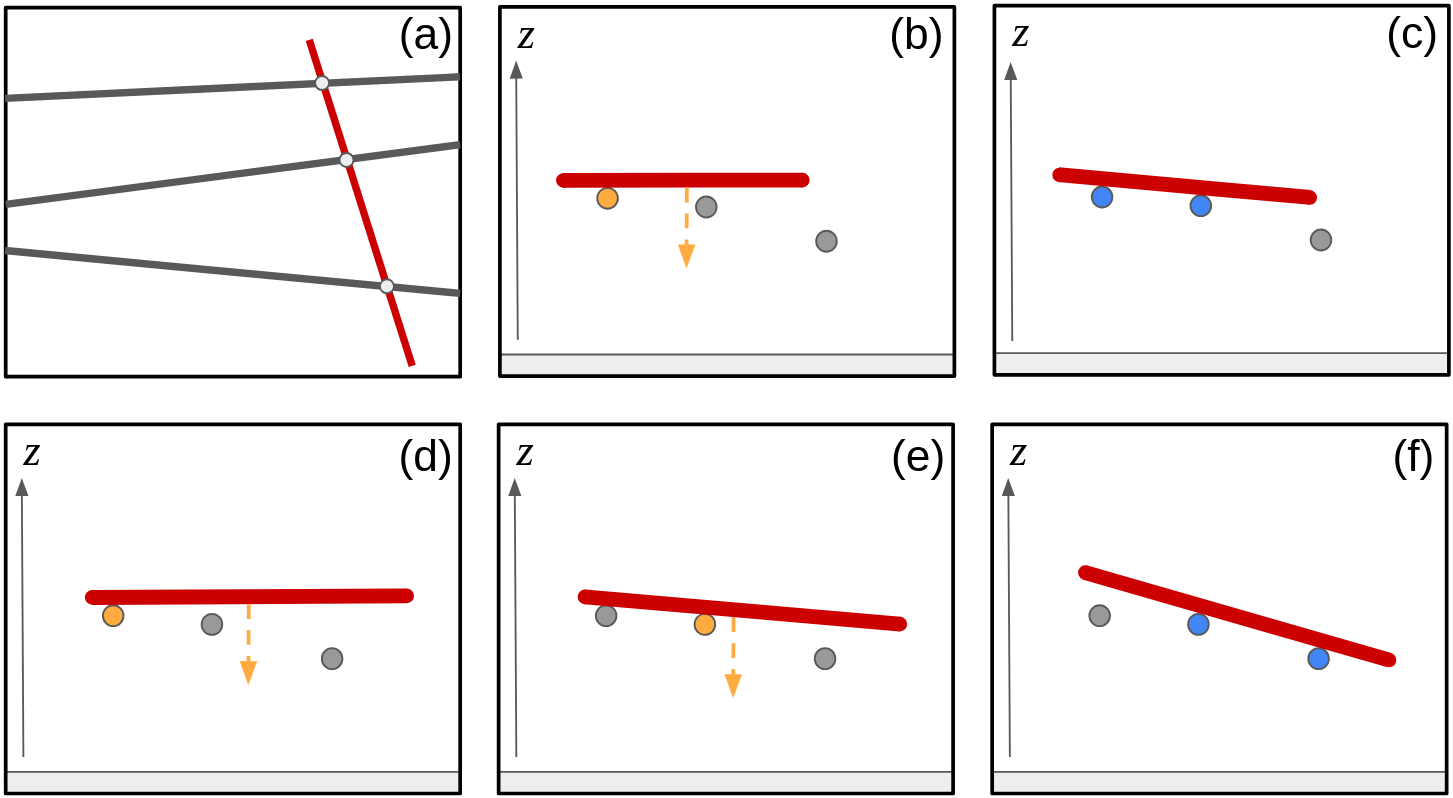}
    \caption{Schematic of the Q3D settling algorithm for two cases. Throughout, orange marks the support currently acting as pivot, blue a confirmed contact, and grey an inactive or lifted crossing; dashed downward arrows indicate the centre of mass of the deposited wire. In the top row, the newly deposited wire settles on the first two encountered wires: (a)~In projection, the deposited wire (red, running from top to bottom) crosses three existing wires (grey, running left to right); each crossing is a candidate contact (open circles). (b)~In side view, the wire first rests on the highest support (orange, far left), which acts as a pivot; with its centre of mass offset from the pivot it tilts toward the heavy side. (c)~The wire settles onto a second support, now resting on two contacts (blue); the remaining crossing (far right) is lifted out of contact and discarded. In the bottom row, the first two wires are insufficient to determine the final resting place of the wire: (d)~When the resulting support pair does not straddle the centre of mass, the configuration is unstable; (e)~the wire re-pivots on the inner support and tilts again until it rests stably with the centre of mass between the two supports~(f).}
    \label{fig:schematic}
\end{figure*}

If one or more crossings exist, the wire initially contacts the highest available support, which acts as a pivot. The wire then tilts under gravity toward its heavy side (i.e.\ the side on which the centre of mass lies, determined by whether the pivot point is to the left or right of the wire midpoint along its length). As the wire rotates, it encounters either a secondary support --- another existing wire on the descending side --- or the substrate. The first obstacle encountered arrests the rotation and defines the wire's final resting configuration, supported at two points.

When the secondary support does not straddle the wire's centre of mass, the resulting two-point configuration is unstable, because the centre of mass lies outside the support span. In this case, the wire re-pivots on the inner support (the one closer to the midpoint) and the tilting procedure repeats. Since each iteration moves the pivot closer to the wire's centre, this process converges in at most a few steps. A schematic of a single iteration of this settling can be seen in Figure~\ref{fig:schematic}.

Once the wire has settled, its $z$-coordinate varies linearly between its two endpoints. The final step is to determine which of the projected crossings correspond to actual physical contacts. A crossing qualifies as a contact only if the settled wire's height at that point matches the support height to within a numerical tolerance. Crossings where the wire has been lifted away by the stacking geometry are discarded. This is the central mechanism by which Q3D stacking reduces network connectivity relative to the 2D model: many projected intersections no longer correspond to physical contacts.

In the 2D simulation, each boundary is subdivided into unit-length sticks that participate directly in intersection tests. In the Q3D model, we instead treat the left and right boundaries as ideal electrodes: any wire whose $xy$-projection crosses $x = 0$ or $x = L$ is connected to the corresponding electrode regardless of its $z$-height. This choice isolates the effect of stacking on \emph{internal} network connectivity from electrode geometry, and ensures that any change in the percolation threshold relative to the 2D case is attributable solely to the reduced contact density within the bulk of the network. To see why this is the case, consider firstly that the percolation threshold is a bulk property defined in the thermodynamic limit and independent of the boundary condition~\cite{StaufferAharony2018, Ziff1992, HoviAharony1996}. Any difference acts only on wires within $\sim l$ of the electrodes, a fraction $\sim l/L$ that vanishes as $L\to\infty$. Since the thresholds are obtained by extrapolation (Eq.~\eqref{eq:Nc_convergence}), the boundary effect is absorbed into the non-universal correction amplitude $b_0$. Secondly, the ideal electrode is moreover the physically faithful choice and preserves the 2D boundary connectivity: in the $xy$-projection, the widthless 2D boundary sticks connect to any wire crossing the edge, and the ideal electrode reproduces exactly this rule regardless of $z$-height, mirroring a macroscopic metallic contact that grabs any wire reaching it and is not itself deposited. Extending the segmented sticks literally into Q3D would instead place the boundary in the stacking hierarchy, allowing electrode wires to be lifted away --- a potentially unphysical boundary-only perturbation. We retain the segmented boundary in 2D solely to reproduce the setup of Li and Zhang~\cite{LiZhang2009} for validation.

All other aspects of the simulation --- random placement, subcell-based intersection testing, union-find cluster tracking, Poisson convolution, and $N_c$ extraction --- are identical to the 2D procedure described above.

\section{\label{sec:results}Results and Discussion}
\subsection{2D validation}

We first validate the simulation and analysis framework against the known 2D result of Li and Zhang. Approximately $10^6$ realisations were performed for each of the system sizes $L = 32, 36, 40, 48, 64, 128, 256$. This is fewer than the $10^7 - 10^8$ realisations used by Li and Zhang, but sufficient for the present purposes: the Q3D simulations that are the primary focus of this work are significantly more computationally intensive, and using the same realisation count for the 2D validation ensures a like-for-like comparison.

\begin{figure}
\includegraphics[width=1.0\columnwidth]{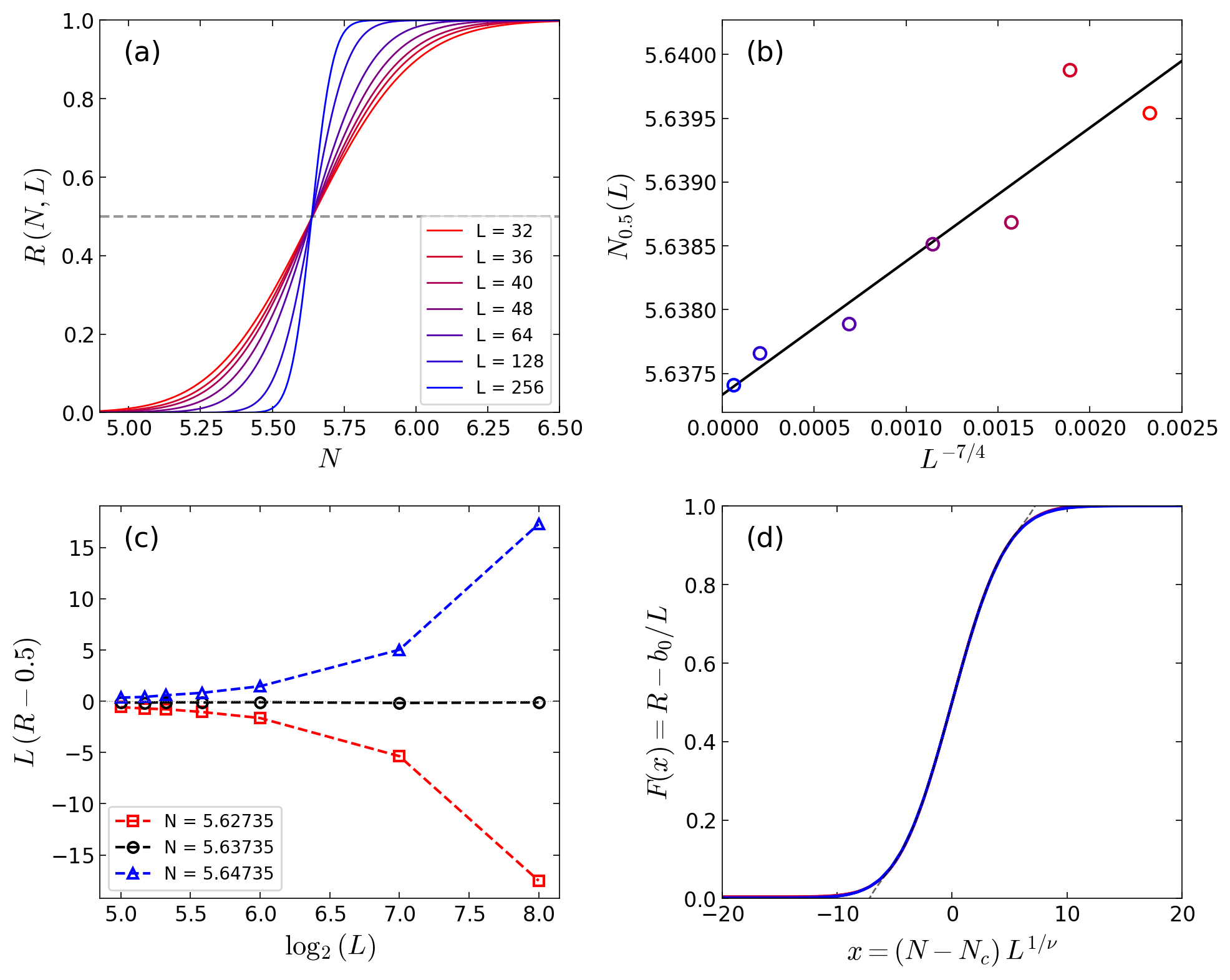}
\caption{\label{fig:2D}Finite-size scaling analysis of the spanning probability for 2D widthless stick percolation.
(a)~Spanning probability $R(N,L)$ as a function of stick density $N$ for system sizes $L = 32$--$256$, with the dashed line indicating $R = 0.5$.
(b)~Finite-size estimates $N_{0.5}(L)$ plotted against $L^{-7/4}$; the solid line shows the linear extrapolation to $L \to \infty$ used to obtain an initial estimate of $N_c$. Points for small system sizes (red) deviate from linearity, while the large-$L$ points (blue) used in the fit are well described by Eq.~\eqref{eq:Nc_convergence}.
(c)~$L(R - 0.5)$ versus $\log_2 L$ for three trial densities near criticality; the flat curve at $N = 5.63735$ identifies the critical density where scale invariance holds.
(d)~Data collapse onto the universal scaling function $F(x) = R - b_0/L$ with $x = (N - N_c)\,L^{1/\nu}$; the dashed line shows the polynomial fit to Eq.~\eqref{eq:polynomial}. The collapse confirms consistency with 2D random percolation universality and serves as a validation of the simulation and fitting methodology against the results of Li and Zhang~\cite{LiZhang2009}.}
\end{figure}

Figure~\ref{fig:2D}(a) shows the spanning probability $R(N, L)$ obtained by Poisson convolution of the raw spanning data. The curves steepen with increasing $L$ and intersect near $R = 0.5$, consistent with the expected finite-size scaling behaviour.

To extract $N_c$  panel~(b) plots $N_{0.5}(L)$ against $L^{-1-1/\nu} = L^{-7/4}$. A linear fit to all system sizes yields $N_c = 5.63735 \pm 0.00001$ as the $L \to \infty$ intercept. This value lies slightly above the Li and Zhang result of $N_c = 5.63726 \pm 0.00002$; the discrepancy of approximately $10^{-4}$ is consistent with the reduced number of realisations used here.

Panel~(c) illustrates the sensitivity of the analysis to the value of $N_c$. The quantity $L(R - 0.5)$, evaluated at three trial densities separated by $\pm 0.01$ from our best estimate, is plotted against $\log_2 L$. At the critical density $N = N_c$, this quantity should remain bounded as $L$ increases, while deviations above or below $N_c$ produce unbounded growth. The central curve remains flat across all system sizes, confirming the extracted threshold.

With $N_c$ fixed, the scaling function coefficients are obtained by fitting Eqs.~\eqref{eq:scaling}--\eqref{eq:polynomial} to the $R(N, L)$ data. Panel~(d) shows the resulting data collapse: the corrected spanning probability $F(x) = R - b_0/L$, plotted against the scaling variable $x = (N - N_c)L^{1/\nu}$, falls onto a single curve for all system sizes. The polynomial fit (dashed line), performed over the range $|x| \leq 5$, yields $K_3 = -1.053 \pm 0.003$ and $K_5 = 0.777 \pm 0.012$, in excellent agreement with the Li and Zhang values of $-1.055 \pm 0.002$ and $0.783 \pm 0.004$ respectively. As an independent check, the simultaneous free-parameter fit returns $N_c$ values differing by less than $5 \times 10^{-6}$ from the two-step estimate. The full set of parameters is reported in Table~\ref{tab:results}. The adjusted $R^2$ of $0.999990$ confirms the quality of the fit.

\subsection{Q3D results}

We apply the same procedure to the Q3D deposition model, using the same set of system sizes and number of realisations per size. Here, we use $d/l = 10^{-3}$.

\begin{figure}
\includegraphics[width=1.0\columnwidth]{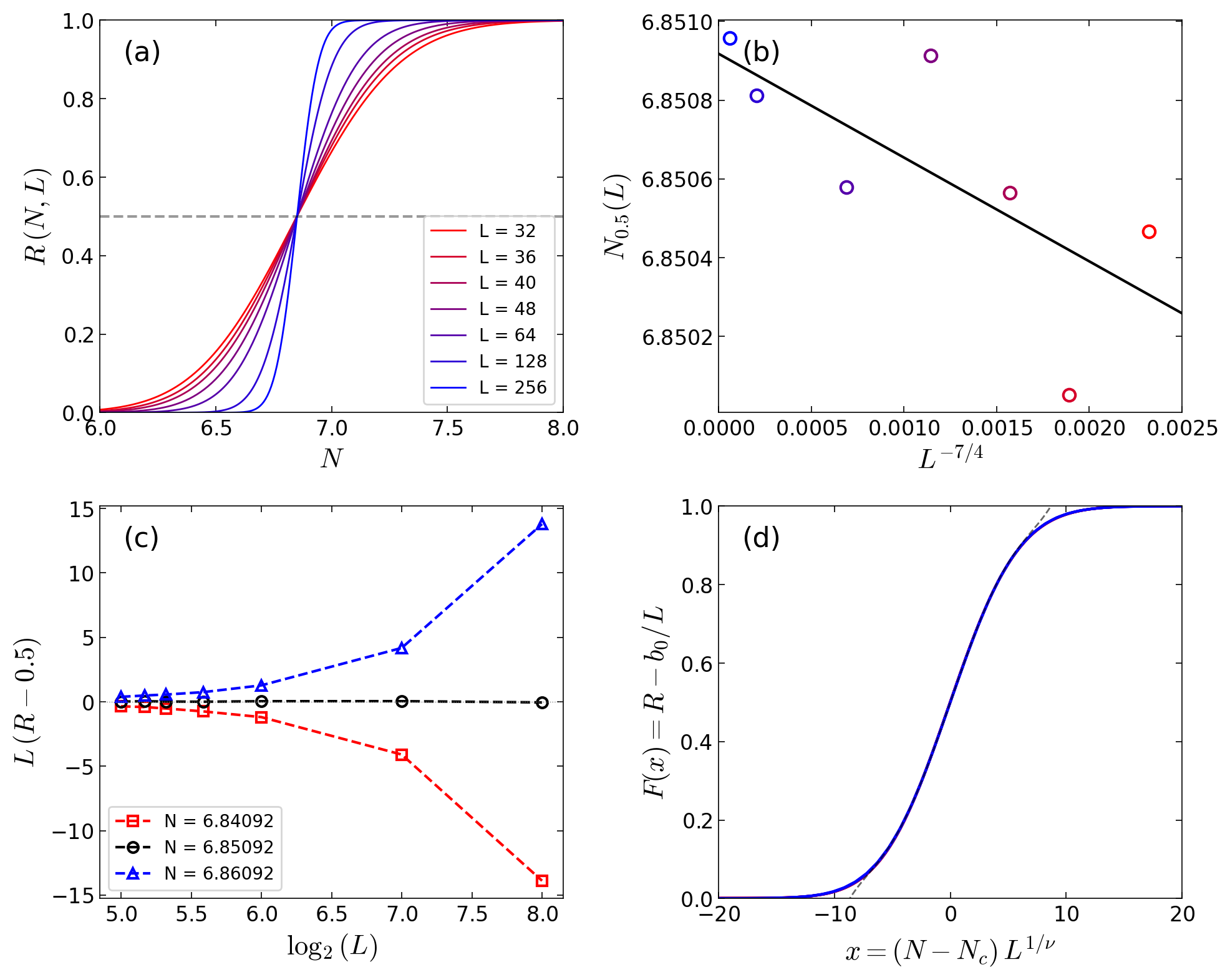}
\caption{\label{fig:Q3D}Finite-size scaling analysis of the spanning probability for quasi-3D nanowire networks.
(a)~Crossing probability $R(N,L)$ as a function of wire density $N$ for system sizes $L = 32$--$256$, with the dashed line indicating $R = 0.5$. (b)~Finite-size estimates $N_{0.5}(L)$ plotted against $L^{-7/4}$; the solid line shows the linear extrapolation to $L \to \infty$ used to obtain an initial estimate of $N_c$. (c)~$L(R - 0.5)$ versus $\log_2 L$ for three trial densities near criticality; the flat curve at $N = 6.85092$ identifies the critical density where scale invariance holds. (d)~Data collapse onto the universal scaling function $F(x) = R - b_0/L$ with $x = (N - N_c)\,L^{1/\nu}$, confirming consistency with 2D random percolation universality. The dashed line shows the polynomial fit to the scaling function.}
\end{figure}

Figure~\ref{fig:Q3D}(a) shows the spanning probability curves for the Q3D system. The qualitative behaviour mirrors the 2D case---curves steepen with $L$ and cross near $R = 0.5$---but the crossing region is shifted to higher density, near $N \approx 6.85$.

The extrapolation in panel~(b) yields $N_c = 6.85092 \pm 0.00001$, approximately $21.5\%$ above the 2D threshold. As in the 2D case, the two smallest system sizes show deviations and we note that the scatter in panel~(b) is somewhat larger than in the 2D case. The sensitivity analysis in panel~(c) again confirms the extracted threshold: the central curve remains flat while the perturbed curves diverge.

Despite the substantial shift in $N_c$, the Q3D spanning data collapse onto a universal scaling function that is visually indistinguishable from the 2D case. The polynomial fit yields $K_3 = -1.075 \pm 0.009$ and $K_5 = 0.885 \pm 0.049$. The $K_3$ value is consistent with the 2D universal values, while $K_5$ shows a larger departure. However, this coefficient is known to exhibit significant scatter across systems and methods: lattice models give $K_5 \approx 1$, Li and Zhang's continuum result is $0.783$, and our 2D value sits slightly below that. The Q3D value of $0.885$ falls within this existing range of variation and does not, by itself, indicate a departure from the 2D universality class. The simultaneous fit again agrees with the two-step method to within $5 \times 10^{-6}$, and the adjusted $R^2$ of $0.999978$ again confirms excellent fit quality.

The nonuniversal parameters do differ between the two models, as expected. The metric factor $A = a_1 = 0.085048 \pm 0.000042$ is approximately $20\%$ smaller than the 2D value of $0.107034 \pm 0.000033$, reflecting the reduced sensitivity of the spanning probability to density changes near threshold: because vertical stacking eliminates a fraction of projected crossings, each additional wire contributes fewer new contacts on average, and the transition from non-spanning to spanning is correspondingly less steep in density. The correction-to-scaling amplitude $b_0$ is also substantially smaller in magnitude ($\sim -0.03577$ vs.\ $\sim -0.155$), indicating that the leading finite-size boundary correction is weaker in the Q3D system. This reflects the use of ideal electrodes, which connect to any wire crossing the boundary regardless of $z$-height, thereby reducing the geometric mismatch between bulk and boundary that drives the $b_0/L$ correction in the 2D case with segmented boundary sticks.

The physical origin of the elevated Q3D threshold is ultimately the reduced connectivity introduced by vertical stacking. In the 2D model, every projected crossing registers as a contact; in the Q3D model, a wire resting on top of previous deposits is lifted away from wires it would otherwise have crossed. Fewer contacts per wire means a higher density is required to build a spanning cluster. That the increase is roughly $21.5\%$ reflects the fact that deposition onto a flat substrate produces relatively modest vertical separation at densities near threshold, where the network is sparse.

\begin{table*}
\caption{\label{tab:results}
Finite-size scaling parameters from the joint fit for the 2D and Q3D stick percolation models. The Li \& Zhang reference values are from Ref.~\cite{LiZhang2009}. The two-step and joint-fit estimates of $N_c$ agree to within $5 \times 10^{-6}$ for both models. Note that Li and Zhang do not give an exact value for $b_0$.}
\begin{ruledtabular}
\begin{tabular}{lddd}
             & \multicolumn{1}{c}{2D} & \multicolumn{1}{c}{Q3D} & \multicolumn{1}{c}{Li \& Zhang (2D)} \\
\hline
$N_c$        & 5.637348(10)        & 6.850923(14)      & 5.63726(2)  \\
$a_1$        & 0.107034(33)      & 0.085048(42)    & 0.106910(9)   \\
$a_3$        & -0.001291(5)      & -0.000661(7)    & -0.001289(2)   \\
$a_5$        & 0.000011(0)       & 0.000004(1)    & 0.00001093(5)   \\
$K_3$        & -1.053(3)            & -1.075(9)     & -1.055(2)   \\
$K_5$        & 0.777(12)             & 0.885(49)       & 0.783(4)    \\
$b_0$        & -0.15455(122)     & -0.03577(155)     & -0.107   \\
$R^2_{\mathrm{adj}}$ & 0.999990 & 0.999978 & 0.999993 \\
\end{tabular}
\end{ruledtabular}
\end{table*}

The Q3D model introduces an additional dimensionless parameter, the diameter-to-length ratio $d/l$. We investigate values spanning several orders of magnitude ($d/l = 10^{-3}$, $10^{-2}$, $10^{-1}$), corresponding to wires ranging from extremely slender to moderately thick relative to their length. For silver nanowires typically used in experiments~\cite{DanielsBrown2021}, diameters are of order $20$--$100\;\mathrm{nm}$ with lengths of several micrometres, giving $d/l \sim 10^{-2}$, which sits comfortably within our range.

Beyond the diameter-to-length ratio, two further microstructural choices are known to influence the percolation threshold in 2D stick systems: the orientational distribution of the wires and dispersity in their length. Both are non-universal: they shift $N_c$ but leave the universality class intact. Macroscopic anisotropy raises the threshold, diverging as the wires approach full alignment (parallel zero-width sticks never cross), whereas length dispersity lowers it, as occasional long wires seed a spanning cluster at reduced number density \cite{BalbergBinenbaumAnderson1983, TarasevichEserkepov2018}. In addition, binary short/long mixtures can show non-monotonic composition dependence \cite{Chaterjee2014, Finner2018}. In the present Q3D model these effects could modulate the threshold as in 2D, with the anisotropy being compounded by the fact that aligned wires already cross less often in projection. A quantitative characterisation across these distributions is left to future work.

One might think that the 2D result is recovered as $d/l \to 0$. However, the contact topology is in fact independent of $d/l$: all vertical coordinates in the system scale with $d$, so the set of crossings that survive the settlement process is determined entirely by the $xy$-geometry and the deposition order. We confirm this numerically, finding no measurable change in $N_c$ across several orders of magnitude in $d/l$. The 2D model is therefore not recovered in the limit $d/l \to 0$; it corresponds instead to the qualitatively different assumption that wires may freely interpenetrate, so that all projected crossings register as contacts regardless of deposition sequence. We note that while $d/l$ does not affect the results, it does control the validity of the model: the settlement algorithm assumes slender rods whose contacts are pointlike. For large $d/l$, effects such as angle-dependent contact geometry, finite wire footprint on the substrate, and rolling would become significant, and a more detailed mechanical model would be required. For the nanowires of experimental interest ($d/l \sim 10^{-2}$), the slender-rod approximation is well justified.

\section{Conclusion}

We have extended the universal finite-size scaling framework for continuum stick percolation from two dimensions to quasi-three-dimensional systems in which wires of finite diameter stack vertically under sequential deposition. The percolation threshold for the Q3D model is $N_c l^2 = 6.850923 \pm 0.000014$, approximately $21.5\%$ above the established 2D value of $5.6373$, reflecting the reduced contact density caused by vertical stacking. The threshold is independent of the wire diameter-to-length ratio $d/l$ over at least three orders of magnitude ($10^{-3}$ to $10^{-1}$), consistent with the scale invariance of the contact topology under sequential deposition.

The spanning probability data for the Q3D system collapse onto the same universal finite-size scaling function as the 2D continuum and lattice models, with universal coefficients consistent with known 2D values. Together with the planar geometry of the connectivity transition (vertical stacking alters which crossings register as contacts but not necessarily the dimensionality of the underlying connectivity problem), this is consistent with Q3D stick percolation belonging to the two-dimensional random percolation universality class. We stress that our analysis assumes $\nu=4/3$ rather than determining it; the collapse confirms consistency with this value but is not an independent measurement of the correlation-length exponent. A definitive assignment of the universality class would require independent determination of the critical exponents, a subject of potential further work. The nonuniversal parameters ($a_1$ and $b_0$) differ between the two models, as discussed in Section~IV, but these do not affect the universality of the scaling function.

These results have practical implications for the modeling of nanowire networks. Device simulations based on the 2D percolation threshold will underestimate the wire density required for electrical connectivity by approximately $21.5\%$, with further consequences for predicted conductivity scaling above threshold~\cite{TarasevichEserkepov2026}. For neuromorphic computing applications, where network dynamics near the percolation transition are of particular interest~\cite{Stieg2012,Mallinson2019}, the correct identification of the critical density is essential for tuning devices to operate at the edge of criticality. Future work could extend this analysis to anisotropic deposition, where preferential wire alignment is expected to further alter the percolation threshold, and to systems with finite junction resistance, where the interplay between contact topology and electrical transport near threshold remains unexplored in the Q3D geometry.

\begin{acknowledgments}
We wish to acknowledge the support of the Accelerate Programme for Scientific Discovery. Thank you to Simon Brown for useful discussions early on in this work. This work was performed in part using resources provided by the Cambridge Service for Data Driven Discovery (CSD3) operated by the University of Cambridge Research Computing Service (www.csd3.cam.ac.uk), provided by Dell EMC and Intel using Tier-2 funding from the Engineering and Physical Sciences Research Council (capital grant EP/T022159/1), and DiRAC funding from the Science and Technology Facilities Council (www.dirac.ac.uk). Additional components of this work were performed with resources granted by RunPod.
\end{acknowledgments}


\bibliography{apssamp}

\end{document}